\documentclass[useAMS,usenatbib]{mn2e}
\usepackage{graphicx}
\usepackage{amsmath}
\usepackage{tabularx}
\usepackage{hyperref}

\usepackage{times}
\usepackage{amssymb,amsmath}

\def\lsim{ \lower .75ex\hbox{$\sim$} \llap{\raise .27ex \hbox{$<$}} }
\def\gsim{ \lower .75ex \hbox{$\sim$} \llap{\raise .27ex \hbox{$>$}} }

\newcommand{\bi}{\begin{itemize}}
\newcommand{\ei}{\end{itemize}}

\title[A stochastic model for blazar variability] 
{On the distribution of fluxes of gamma-ray blazars: hints for a stochastic process?} 

\author[Tavecchio et al.]
{F. Tavecchio$^1$\thanks{E--mail: fabrizio.tavecchio@brera.inaf.it}, G. Bonnoli$^{2,3}$ and G. Galanti$^1$ 
\\
$^1$INAF -- Osservatorio Astronomico di Brera, via E. Bianchi 46, I--23807 Merate, Italy\\
$^2$Dipartimento di Scienze Fisiche, della Terra e dell'Ambiente - Universit\`{a} degli Studi di Siena, via Roma 56, I--53100 Siena, Italy\\
$^3$INFN - Sezione di Pisa, largo B. Pontecorvo, 3 I--56127 Pisa, Italy\\
}

\begin{document}



\maketitle

\begin{abstract} 
We examine a model for the observed temporal variability of powerful blazars in the $\gamma$-ray band in which the dynamics is described in terms of a stochastic differential equation, including the contribution of a deterministic drift and a stochastic term. The form of the equation is motivated by the current astrophysical framework, accepting that jets are powered through the extraction of the rotational energy of the central supermassive black hole mediated by magnetic fields supported by a so-called \emph{magnetically arrested} accretion disk. We apply the model to the $\gamma$-ray light curves of several bright blazars and we infer the parameters suitable to describe them. In particular, we examine the differential distribution of fluxes ($dN/dF_{\gamma}$) and we show that the predicted probability density function for the assumed stochastic equation naturally reproduces the observed power law shape at large fluxes $dN/dF_{\gamma} \propto F_{\gamma}^{-\alpha}$ with $\alpha>2$. 
\end{abstract}

\begin{keywords} galaxies: jets --- radiation mechanisms: non-thermal ---   Gamma--rays: galaxies 
\end{keywords}

\section{Introduction}

Blazars are the most luminous persistent sources in the Universe (e.g. Romero et al. 2017). In their core a supermassive black hole accretes matter from the surrounding host galaxy and part of the released gravitational energy is conveyed to a collimated relativistic (typical Lorentz factor $\approx 10$) outflow of plasma whose axis points close to the Earth (Blandford \& Rees 1978). In this geometry, relativistic effects greatly enhance the observed luminonosity of the non-thermal radiation produced by ultra-relativistic particles energized in the flow, so that this component often outshines the thermal contribution from the nucleus. The observed emission is characterized by a spectral energy distribution (SED) with two well defined bumps (e.g. Ghisellini et al. 2017). The low energy component is associated to synchrotron radiation from relativistic electrons, while the high-energy hump is likely produced through inverse Compton emission by the same electrons (e.g., Maraschi et al. 1992), although contribution from hadronic processes cannot be excluded (e.g. Boettcher et al. 2013, Cerruti et al. 2015). 

Violent variability, both in amplitude and time scale, is one of the defining properties of blazars. Variability is observed to be more extreme in the $\gamma$-ray band, where flux variations by several orders of magnitude (e.g. Bonnoli et al. 2011, Ghirlanda et al. 2011) and flares lasting few minutes (e.g. Aharonian et al. 2007, Aleksi{\'c} et al. 2011,2014) are often recorded. The observed flux variability can be used as an extremely powerful tool to test emission models, to constrain in size and locate the emission region(s) and to investigate particle acceleration processes (e.g. Blandford et al. 2019). However, despite extensive investigation, it is far from clear if variability is mainly connected to the physical processes occurring close to the central engine (i.e. a time-dependent power injection) or, instead, the main driver is the variable rate of jet energy dissipation. Current relativistic magneto-hydro-dynamical (RMHD) simulations (e.g. Tchekhovskoy et al. 2011, White et al. 2019) agree that jets are most efficiently fed when the accretion flow reaches the so called {\it magnetically arrested disk} (MAD) condition and the rate of energy extraction (occurring mainly via the Blandford \& Znajek 1977 process) is naturally modulated, tracking the fluctuations of the magnetic flux in the innermost regions of the disk.

Operational since August 2008, the Large Area Telescope (LAT) on board the {\it Fermi} satellite (Atwood et al. 2009) has accumulated an unprecedented wealth of blazar data. In particular, due to its operative mode, LAT provides intensive monitoring of sources in every region of the sky, making it possible to obtain densely sampled high-energy lightcurves of unprecedented detail, extension and duty-cycle (e.g. Tavecchio et al. 2010, Abdo et al. 2010, Nalewajko 2013). Recently, Meyer, Scargle and Blandford (2019, hereafter MSB19) reported a detailed analysis of variability of bright $\gamma$-ray flat spectrum radio quasars (FSRQ) based on LAT lightcurves. In particular, they were able to derive statistically rich flux distributions, $dN/dF_{\gamma}$, providing information on the relative frequency of states with different fluxes. These distributions were studied already at the beginning of the {\it Fermi} mission (e.g. Tavecchio et al. 2010), but the longer time span allows MSB19 to have a much clearer description on the underlying shape. In particular, all the six blazars considered share remarkably similar distributions, well described by a broken power law (but also consistent with a log-parabolic shape). Although at low fluxes the statistic does not allow to draw a firm conclusion, at large fluxes the distributions are quite well described by a steep power law, $dN/dF_{\gamma} \propto F_{\gamma}^{-\alpha}$ with $\alpha>2$ . Such a well defined flux distribution for all considered sources clearly calls for an explanation. A possibility, mentioned by MSB19, is self-organized criticality, that naturally predicts power law frequency distributions, as observed in solar flares (see, e.g., Aschwanden et al. 2016). Remarkably, this behavior is at odds with the log-normality observed in other bands in some blazars (e.g. Giebels \& Degrange 2009) and that points to different interpretations, namely independence of the physical parameters controlling the observed emission. 

In this paper we explore a possible description of blazar variability in terms of a stochastic process following an underlying stochastic differential equation (SDE). In brief, we postulate that variations are driven by the interplay between a deterministic process trying to maintain an equilibrium state and a stochastic "noise" continuously pushing the system out of stability. We identify the innermost region of the accretion disk (where the jet is launched) as the most natural location for such a process and the magnetic field as the main actor. While in the past some papers describing quasar and AGN variability in terms of stochastic processes described by SDE already appeared (e.g. Vio et al. 1992, 1993, Kelly et al. 2009; Kozlowski et al. 2010), this topic has received limited attention for blazars so far. Recently, Sobolewska et al. (2014) considered SDE to model $\gamma$-ray light curves of blazars, but they considered a stochastic term (describing the classical Ornstein-Uhlenbeck process) that, as we discuss hereafter, we consider unsuitable to reproduce blazar dynamics. Moreover they focused the analysis on the power spectrum of blazar light curves. Our approach is instead to consider a SDE tailored on the specific processes that we believe regulate the central engine.

The paper is structured as follows. In sect. 2 we describe the underlying astrophysical scenario and the stochastic differential equation based on it. In sect 3. we apply the model to the observational data, in particular to reproduce the flux distribution of the blazars. Eventually in sect. 4 we discuss our results.

\section{A physically inspired stochastic model for variability}

Systems whose dynamics is the result of both deterministic and random contributions, are best modeled in terms of stochastic differential equations (SDE). In the general form (for simplicity we assume a one-dimensional system) a SDE can be written as:
\begin{equation}
dX = f(X,t)dt + \Sigma(X,t)dW_t
\label{eq:gensde}
\end{equation}
where $X(t)$ is the (stochastic) variable whose time evolution we intend to describe, $f(X,t)$ is a function, usually called {\it drift}, modelling the deterministic ``force" acting on the system and $\Sigma(X,t)$ is the function specifying the random term, driven by the standard stochastic Wiener process $W_t$ (often called brownian motion, since it is strictly related to its modeling). For a complete view we refer the reader to the numerous  textbooks dedicated to the subject (e.g. Allen 2007).

Assuming the measured $\gamma$-ray flux as the time-dependent variable, $X(t)\equiv F_{\rm \gamma}(t)$, our aim is to determine the functions $f$ and $\Sigma$ taking inspiration from the current knowledge of the disk-jet system in blazars. 

\subsection{The astrophysical scenario}

The scenario that we would like to explore assumes that the modulation of the $\gamma$-ray flux is mainly (although perhaps not completely) driven by variations of the power injected at the base of the jet.

State-of-the-art numerical simulations (e.g. McKinney et al. 2012, Tchekhovskoy et al. 2011) support the view that powerful jets develop in systems in which the accretion flow occurs in the so-called {\it magnetically arrested disk} (MAD) regime (e.g. Narayan et al. 2003, McKinney et al. 2012, Tchekhovskoy 2015). Put simply, in this conditions the system reaches a self-regulated state where the magnetic field carried by the accreting plasma accumulates close to the disk inner edge and reaches a maximum value dictated by the condition that its pressure equals the pressure of the falling matter. If the magnetic pressure exceeds this limit, the accretion of gas is halted and, without the supply of fresh field, the magnetic pressure decreases, until the plasma is allowed to restart accretion. This feedback mechanism is therefore able to maintain the maximum possible magnetic field pressure (or, equivalently, energy density) close to the black hole horizon. 

The value of the magnetic field is a critical parameter dictating the power that the system is able to inject into the jet. Simulations support the view that the power is extracted from the system through the Blandford-Znajek (1977) mechanism, in which the power goes with the square of the magnetic flux $\phi_B$ close to the BH horizon which, in turn, is proportional to the magnetic field in the same region $B$, $\phi_B\propto B$. We therefore assume that the power of the jet is proportional to the energy density of the magnetic field $U_B=B^2/8\pi$. 
In FSRQ under study here the total radiated luminosity is dominated by the gamma-ray component from the external Compton process (e.g. Ghisellini et al. 2010). If we assume a constant radiative efficiency for the jet (i.e. a constant ratio between the jet luminosity and its power), the postulated linear dependence between the energy density of the magnetic fields in the disk and the jet power naturally translates into a linear dependence between  $U_B$ and the observed luminosity or flux, i.e. $F_{\gamma}\propto B^2$ (see also Ghisellini et al. 2010).

While the mechanism described above allows the system to maintain a stable equilibrium characterized by a given magnetic field $B$ close to the horizon, several processes and instabilities likely intervene and perturbate it. For instance, reconnection of field lines is likely to occur locally in the flow, leading to stochastic dissipation of magnetic energy (e.g. Lazarian et al. 2016). On the other hand, in the conditions characterizing the BH vicinity, the plasma is likely to support turbulent motion that can locally amplify the field through dynamo processes (e.g. Arlt \& R{\"u}diger 1999). All these phenomena (dissipation through reconnection, amplification through dynamo) can be thought as a random ``noise" continuously perturbing the equilibrium state. Although a complete treatment should consider  the spatial distribution of the perturbations, in our heuristic approach we treat them as a spatially averaged stochastic term.

\subsection{The stochastic equation}

The accretion-ejection system is of course rather complex and  characterized by several concurrent processes acting at different spatial and temporal scales. We do not pretend to fully catch this complexity reducing the number of degrees of freedom to few variables. Instead our aim is to try to grasp the gross features of the dynamics exploiting a very simplified view of the real situation.

Within this approach, the dynamics can be thought as a combination of a deterministic process trying to keep the equilibrium value of the magnetic energy density against random noise continuously disturbing it. Of course we have some freedom to select the specific expressions for the drift and the stochastic terms. The chain of arguments discussed above motivates us to associate the observed flux to the magnetic energy density close to the BH, $B^2$. We therefore consider a SDE for the magnetic energy density and we identify it with the stochastic variable $X$. In virtue of the direct dependency, the dynamics of the magnetic energy density can be directly translated to that of the recorded flux.

\begin{figure}
 \hspace*{-1.2truecm}
 \vspace*{-0.6truecm}
 \includegraphics[width=0.6\textwidth]{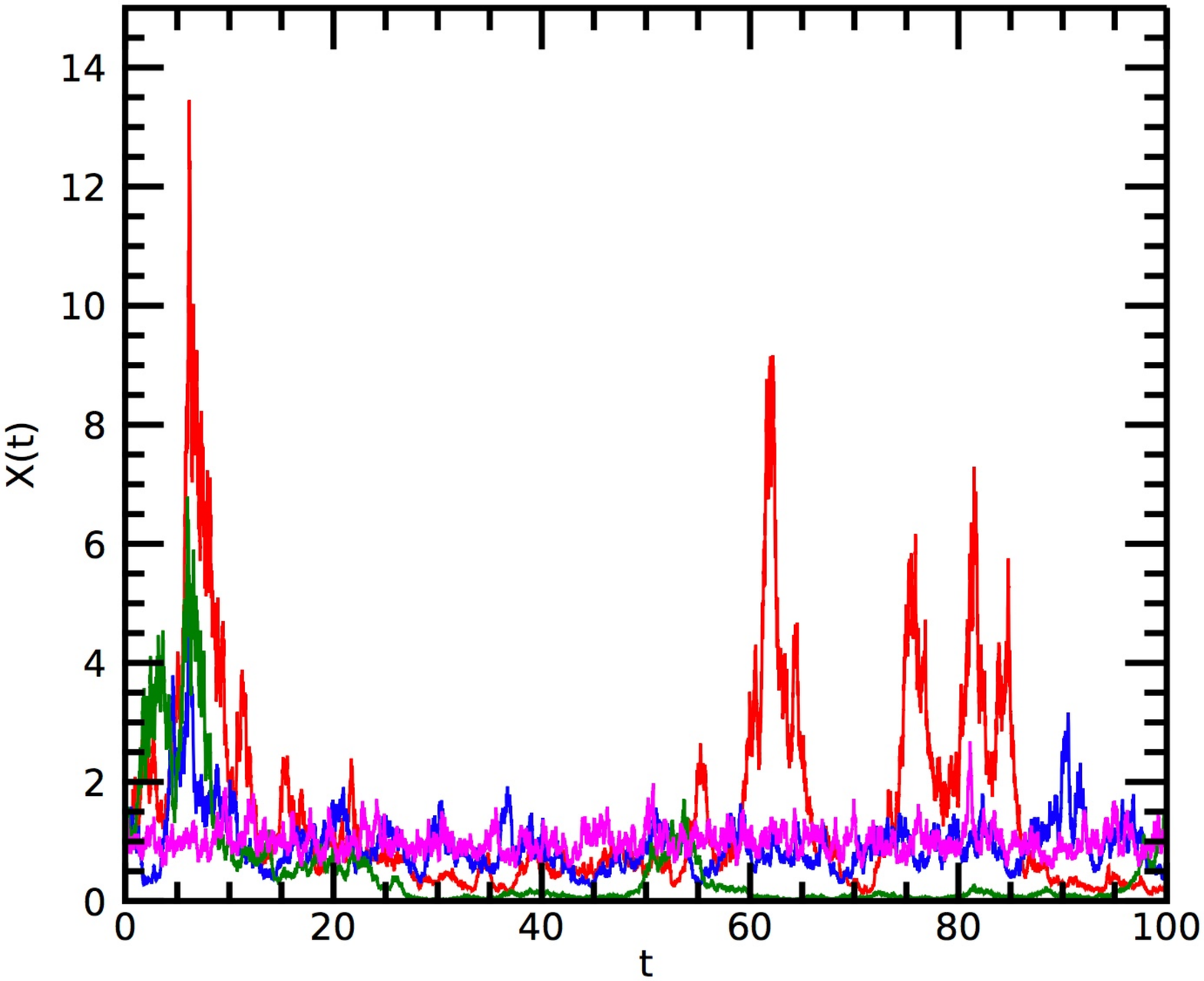}\\
 \vspace*{-0.truecm}
 \caption{Light curves simulated by numerically solving  Eq. \ref{eq:sde}. In all cases we fix $\mu=1$, $\sigma=0.5$. The different curves are calculated for $\theta=0.01, 0.1, 0.5$ and 3 (green, red, blue and magenta). The initial condition is $X_0=1$.}
 \label{fig:lc}
\end{figure}

The simplest expression for $f(X)$ (see eq. \ref{eq:gensde}) which models the tendency to reach an equilibrium state is a linear combination of the form $f(X)\propto (\mu-X)$, where $\mu$ represent the equilibrium value of $X$, for which $\dot{X}=0$. This specific form for the drift term is justified considering the simple description of the system as evolving under the competition between the magnetic and gravitational forces given by Tchekhovskoy (2015).
For the stochastic term, a possibility (more below)  is that its amplitude is proportional to the level of $X$, $\Sigma(X)\propto X$.
Therefore, the astrophysical scenario motivates us to propose the following SDE:
\begin{equation}
dX = \theta(\mu-X)dt + \sigma X dW_t,
\label{eq:sde}
\end{equation}
specified by the parameters $\theta$ (the inverse of the time scale of the drift term), $\mu$ (equilibrium value for $X$) and $\sigma$ (coefficient of the stochastic term).
Therefore $f(X)=\theta(\mu-X)$ and $\Sigma(X)=\sigma X$. Eq. \ref{eq:sde} describes a quite simple underlying dynamics: the drift term pushes the system toward an equilibrium value $\mu$, while the evolution is disturbed by a random noise whose amplitude is proportional to the actual value of $X$. Hence, high states, characterized by large $X$ will also display the largest fluctuations.

The particular dependence of the stochastic term assumed above can be supported by the consideration that the MHD differential equation for the magnetic field including the resistivity term (e.g. Kulsrud 2005) has a form $dB/dt \propto vB+CB/L^2$, where $v$ is the velocity flow, C a constant and $L$ a characteristic scale length. The key point is that both field amplification by dynamo processes (described by the first term on the right) and diffusion/reconnection effects (depending on the second term) depend {\it linearly} on the field intensity. The stochastic term in our SDE, which in our scenario is linked to field amplification/dissipation processes, is therefore expected to be described by a linear term  on $X$. As we will see later, this specific form of the stochastic term is also suitable to describe the observed shape of the flux distribution.

The parameter $1/\theta$, which has units of time, quantifies the timescale associated to the drift term. Although one expects that this parameter is associated to the typical time on which the magnetic field accumulates in the innermost region of the accretion flow, it is difficult to provide an analytical estimate based on the physics of the accretion flow. An indication of its value, however, can be derived from the results of the MHD simulations. In particular, the simulations reported in Tchekhovskoy  et al. (2011) show that the magnetic flux close to the BH erratically varies around an equilibrium values with approximate timescale of the order of $10^3 r_g/c$, where $r_g=GM/c^2$ is the gravitational radius of the BH of mass $M$. Considering that for FSRQ typical BH masses are of the order of few times $10^8$ $M_{\odot}$ (e.g. Ghisellini \& Tavecchio 2015), the expected variations are expected to occur on a timescale of the order of 20-30 days (i.e. $\theta\sim 0.05$).

The estimate of $\sigma$ is more difficult. In principle, this parameter measures the strength of the stochastic perturbation to the system and should be related to the dynamics of the amplification/dissipation processes. In practice it is hard to provide an estimate of this quantity. The comparison with the observations can thus be used to constrain this parameter and, in principle, the dynamics of the processes.

SDE can be numerically solved with standard methods using discretization techniques closely similar to those adopted for ordinary differential equations. Some (discrete) realizations of Eq. \ref{eq:sde} obtained by using the standard Milstein scheme (e.g. Iacus 2008) are shown in Fig.\ref{fig:lc}. In all cases we fix the equilibrium drift value $\mu=1$ and the random noise parameter $\sigma=0.5$ and vary the intensity of the drift term with $\theta=0.01, 0.1, 0.5$ and 3 (green, red, blue and magenta). The most evident feature of the synthetic light curves is that the system never settles into a steady state but $X(t)$ describes a fluctuating evolution with episodic flares whose amplitude is the largest for $\theta= 0.1$. The largest ``outbursts" are separated by long period of relatively quiescent level in which $X$ fluctuates around $\mu$. We remark that the lightcurves presented here are just few  possible realizations, since the stochasticity of the process does not allow one to derive a unique solution of the SDE.

\begin{figure}
 \hspace*{-1.2truecm}
 \vspace*{-0.6truecm}
 \includegraphics[width=0.6\textwidth]{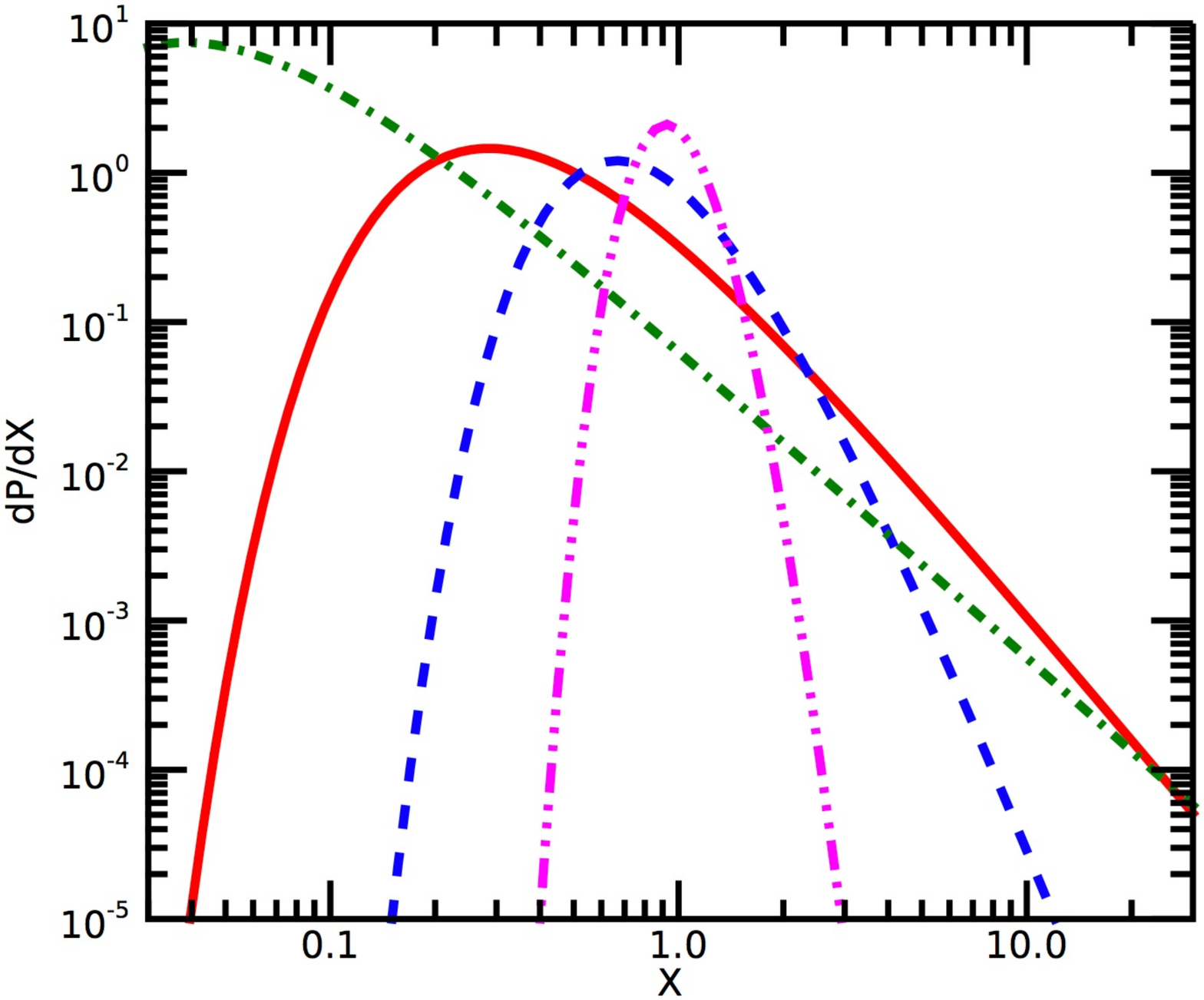}\\
 \vspace*{-0.truecm}
 \caption{Probability density functions corresponding to light curves in Fig.\ref{fig:lc}. In all cases we fix $\mu=1$, $\sigma=0.5$. Curves assume $\theta=0.01, 0.1, 0.5$ and 3 (green, red, blue and magenta), corresponding to $\lambda=0.08, 0.8, 4.0$ and 24.}
 \label{fig:pdf}
\end{figure}

\subsection{Probability density function}

A SDE can be associated to a probability density function (PDF) $p(X,t)$ which provides the relative frequency of the values of the stochastic variable $X(t)$ (see details in e.g. Allen et al. 2007). The time-dependent PDF can be derived from the corresponding Fokker-Planck equation (also known as Kolmogorov equation): 
\begin{equation}
\frac{\partial p(X,t)}{\partial t}+\frac{\partial}{\partial X}[f(X,t)p(X,t)]=\frac{\partial^2}{\partial X^2}\left[ \frac{\Sigma(X,t)^2}{2}p(X,t)\right].
\label{eq:fokker}
\end{equation}
One can easily recognize that the stochastic section plays the role of {\it diffusion} term with an effective diffusion coefficient $D(X,t)=\Sigma(X,t)^2/2$.

Considering a system for a time longer than its relaxation time-scale, it can be described in terms of a stationary state, and the corresponding time-independent PDF is derived by setting $\partial p/\partial t=0$ in the Fokker-Planck equation. Specializing Eq. \ref{eq:fokker} to our case, the steady state PDF, $p(X)$, satisfies the equation:
\begin{equation}
\frac{d}{d X}[\theta(\mu-X)p(X)]-\frac{d^2}{d X^2}\left[ \frac{\sigma^2 X^2}{2} p(X)\right]=0,
\label{eq:fokker2}
\end{equation}
whose solution is (see Appendix A):
\begin{equation}
p(X)=k \frac{e^{-\lambda\mu/X}}{X^{\lambda+2}},
\label{eq:pdf}
\end{equation}
where we have defined $\lambda \equiv 2\theta/\sigma^2$ and $k$ is determined by the normalization condition:
\begin{equation}
\int_0^{\infty}dX \, p(X)=1
\label{eq:fokkerApp3}
\end{equation}
(see also Appendix A).

The PDF calculated for the set of parameters used to simulate the light curves in Fig.\ref{fig:lc} are reported in Fig.\ref{fig:pdf}. The PDF displays a quite simple structure, i.e. it describes a power law with slope $-(\lambda+2)$ above the peak (located at $X_{\rm max}=\lambda\mu/[\lambda+2]$) and an exponential roll-off for $X<X_{\rm max}$. 

\begin{figure}
 \hspace*{-0.18truecm}
 \vspace*{-0.4truecm}
 \includegraphics[width=0.52\textwidth]{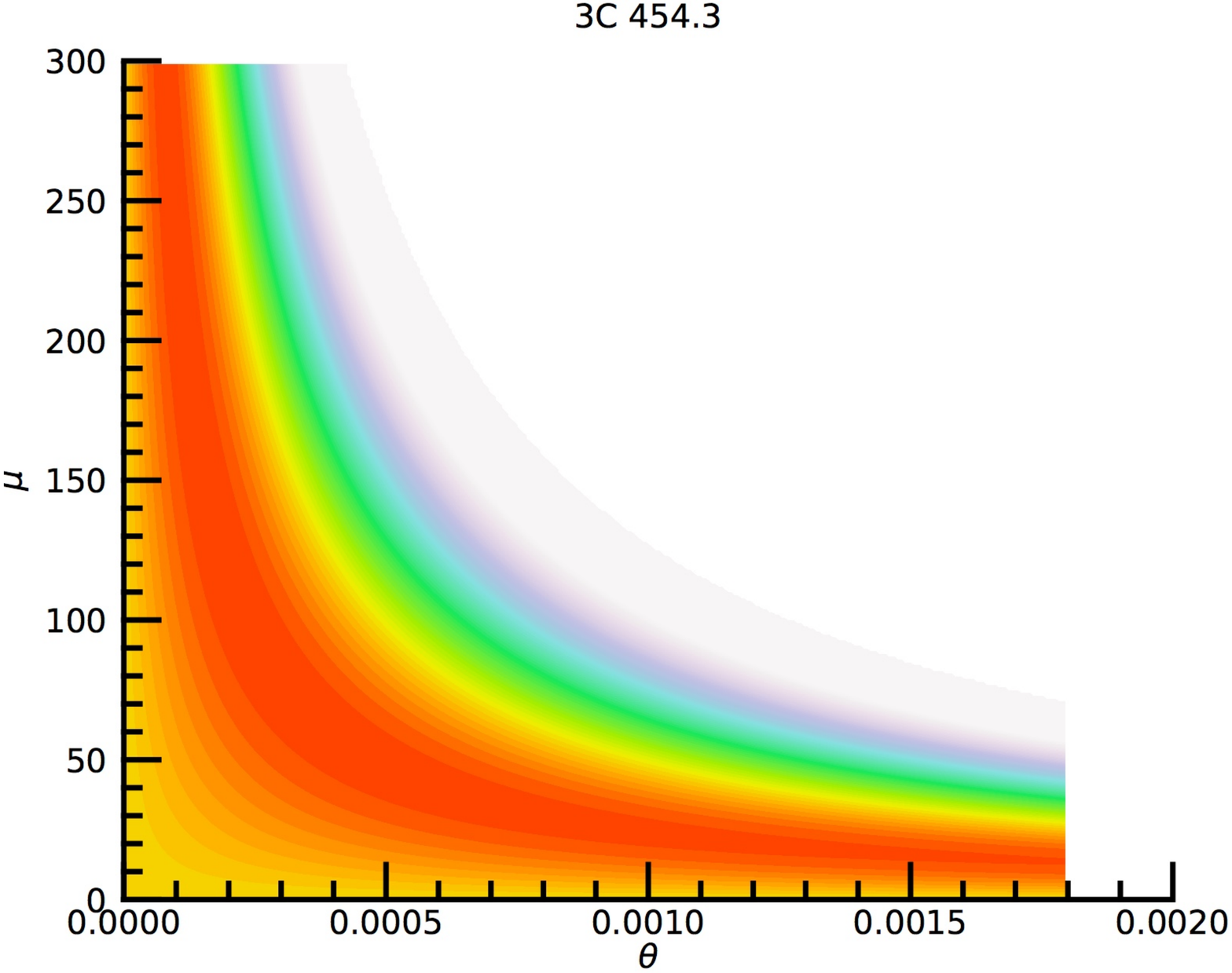}\\
 \vspace*{-0.truecm}
 \caption{Likelihood profile obtained for the parameters $\theta$ and $\mu$ using the lightcurve of 3C 454.3. The shallow maximum (red) extends on a hyperbolic region, showing that the parameters are highly degenerate and no unique solution can be derived.}
 \label{fig:like454}
\end{figure}

The shape of $p(X)$ depends on the value of $\lambda$, i.e. on the relative weight of the drift and stochastic (diffusion) terms. Large $\lambda$, characterizing cases in which the drift overcomes the random noise, are described by narrow PDF centered on $X=\mu$ (magenta line). Decreasing $\lambda$, the importance of stochastic term increases, determining the broadening of the distribution, the hardening of the power law and the shift of the peak to lower $X$ (blue and red curves). The limit $\lambda\to 0$ (describing a system dominated by the stochastic term) is described by a pure power law distribution $p(X)\to k X^{-2}$ (green line). 

We are now in the position to fully appreciate the relevance of the $X$ dependence of the stochastic term of Eq. \ref{eq:sde}. Indeed, without this term the system would describe the standard Ornstein-Uhlenbeck process (describing, for instance, the velocity of a massive particle undergoing Brownian motion under the effect of friction), which is characterized by a stationary {\it gaussian} PDF, clearly not suitable to reproduce the observed $dN/dF_{\gamma}$.

\begin{table*}
\centering
\begin{tabular}{lcccc}
\hline
\hline
Source  & $\sigma$ &$\mu$ & $\theta$ &  $\lambda$\\
\hline
PKS 1222+216 & $0.35\pm0.05$ & 2.1 (1.8-2.6) & 0.04 (0.03-0.05)& 0.15$\pm$0.1 \\
CTA 102 & $0.39\pm0.05$  & -- &-- & 0.2 $\pm$0.15 \\
3C 273 & $0.44\pm0.05$   & 3.6 (2.9-5.5) &0.025 (0.015-0.03)& 0.6$\pm$0.16   \\
3C454.3 & $0.23\pm0.05$  & -- & -- & 0.1 (fixed) \\
PKS 1510-089 & $0.46\pm0.04$   & 6.3 (5.3-7.9) & 0.04 (0.03-0.05)& 0.1 (fixed)  \\
3C279 & $0.44\pm0.04$  & 5.5 (4.0-9.0)& 0.03 (0.015-0.04)& 0.7$\pm$0.1 \\
\hline
\hline
\end{tabular}
\caption{Parameter of the stochastic model derived for the six blazars discussed in the text. Uncertainties are reported at 95\% C.L. The parameter $\mu$ is normalized to $10^{-7}$ cm$^{-2}$ s$^{-1}$.}
\label{tab:param}
\end{table*}

\section{Comparison with blazar light curves}

We apply the model developed in the preceding section to the well sampled lightcurves of six bright FSRQ derived by MSB19\footnote{Lightcurves and flux distributions can be downloaded from: {\tt https://zenodo.org/record/2598791\#.XnYJdW57nE4}}. Specifically, we use the weekly binned lightcurves and we restrict the analysis to bins where excess from the source is statistically significant at the level of $TS>9$, where the test statistics $TS$ (see e.g.  Mattox et al. 1996) is based on the standard likelihood ratio test between a model considering only backgrounds and known field sources  and the one including also a point source for the FSRQ.

Assuming that the dynamics of a system is described by a SDE, standard inference methods allow one to extract the value of the underlying parameters from the observed time series. Methods are based on the generation of a pseudo-likelihood function in which, since an explicit expression for the transition probability cannot be obtained, one inserts a discrete approximation for it using similar schemes developed to solve SDE (see e.g. Allen 2007). 

\begin{figure*}
 \centering
 \hspace*{-1.1truecm}
 \includegraphics[width=1.05\textwidth]{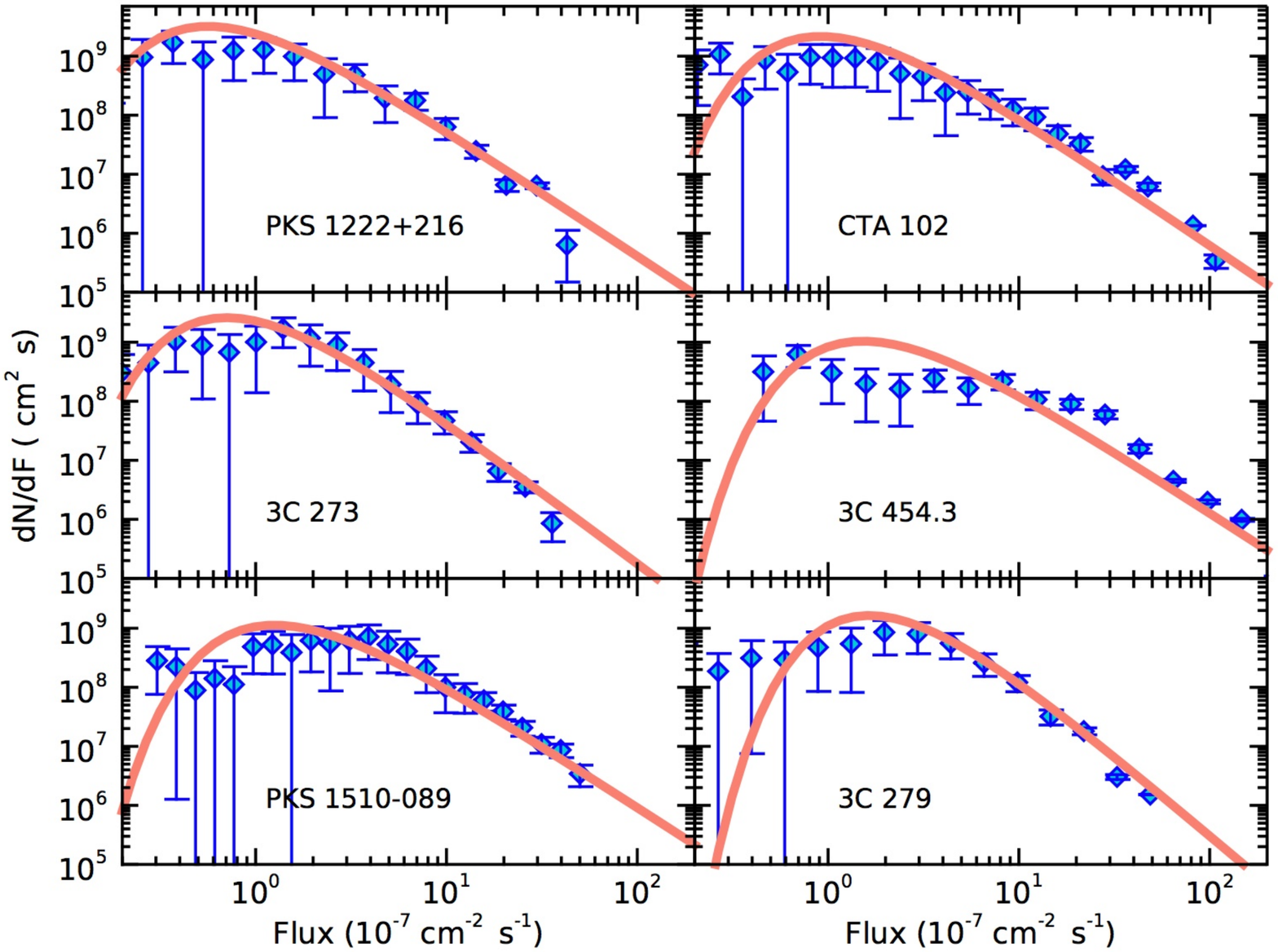}\\
 \vspace*{-0.truecm}
 \caption{Flux distributions for the six blazars considered in our analysis (from MSB19). The red solid curves show the result of the best fit with the probability density function associated to our stochastic equation. See text for details.}
 \label{fig:combo}
\end{figure*}

For our SDE the free parameters are $\sigma$, $\theta$ and $\mu$. The coefficient of the stochastic term is the easiest to estimate. Indeed, if the number $n+1$ of measurements $\{X_i\}$ (with $i=0... n$) is large enough, the maximization of the pseudo-likelihood provides the following expression:
\begin{equation}
\sigma^2\simeq \frac{1}{n}\sum_{i=1}^n\frac{(X_i-X_{i-1})^2} {X_{i-1}^2 (t_i-t_{i-1})}
\end{equation} 
that can be directly used.

The maximization of the likelihood with respect the two other parameters provides two expressions that can be  numerically solved to find estimates of $\theta$ and $\mu$. We report in Table \ref{tab:param} the value of the parameters estimated with the likelihood approach with the corresponding 95\% confidence level uncertainty. For four cases we obtain similar values, namely  $\sigma$  around 0.4, $\theta$ in the range 0.04-0.03 (corresponding to drift timescales of the order of 4 weeks) and $\mu$ between 2 and 6. From these values ($\sigma^2 > \theta$) we conclude that the influence of the stochastic term is important in shaping the observed variability, as also testified by the relatively small values of $\lambda$ derived below.

For two sources, 3C454.3 and CTA 102 in the absence of convergence we were not able to derive a value for $\theta$ and $\mu$. The inspection of the likelihood profiles (fig. \ref{fig:like454}) reveals that the maximum of the pseudo-likelihood traces an extended hyperbolic-like region in the $\mu-\theta$ parameter space, indicating that the two parameters are highly degenerate. 

The comparison between the flux distributions and the expected PDF  provides supplementary information. In fig. \ref{fig:combo} we report the flux distributions of the six blazars derived by MSB19. In all cases the overall shape of the distribution, a power law tail at large fluxes, accompanied by a hardening/plateau at low fluxes, is very similar. As we already mentioned, MSB19 fit these distributions by using a broken power law. However, while the presence of a power law at large fluxes appears a robust feature, the detailed form of the distribution at low fluxes is less clear (the case of 3C454.3 is perhaps the more convincing).

We fit the flux distributions with the PDF obtained in Eq.\ref{eq:pdf}. The resulting fits are shown by the red lines in Fig. \ref{fig:combo}. In all cases the curves satisfactorily reproduce the data. However, for 3C454.3 and PKS 1510-089 the fit does not converge and we are not able to derive the best value for $\lambda$. The curves shown for these two sources have been obtained fixing $\lambda=0.1$. We checked that lower values do not substantially improve the agreement with the data, while for larger values it worsens. 

In the majority of cases the value of $\lambda$ is small ($\lambda\lesssim 0.2$), confirming the prevalence of the stochastic term over the deterministic drift. 3C 279 and 3C 273, instead, show a softer power law requiring a slightly larger $\lambda=0.6-0.7$. The peak of the PDF lies for all sources in the range $0.3-3\times 10^{-7}$ ph cm$^{-2}$ s$^{-1}$.
In any case, the limited statistics does not allow any strong conclusion about possible differences among the sources.

\section{Discussion}

We have proposed a simple model for the variability of blazars exploiting a stochastic differential equation including a deterministic term -- which tends to maintain the system in a stable equilibrium -- and a random noise disturbing it and triggering the variations. The adopted SDE is thought as a rather simplified description of the dynamics of an accretion-jet system in a MAD regime, where the equilibrium is determined by the balance between the repelling magnetic force and the gravitational pull on the accreting material. 

We have assumed that the main parameter controlling the bolometric output of blazar jets is the energy flux (or power), directly linked to magnetic flux threading the black hole horizon. We used this scheme to model well sampled $\gamma$-ray light-curves of six bright FSRQ. For these sources the radiative output is dominated by the $\gamma$-ray component mainly contributing to the LAT band and therefore the $\gamma$-ray lightcurves can be considered good tracers of the bolometric emission and its underlying dynamics. 

The model that we have postulated is able to reproduce in a natural way the shape of the $\gamma$-ray flux distributions, in particular the power law tail at high fluxes, whose slope in our interpretation is determined by the relative weight of the deterministic and the stochastic terms. It is interesting to note here that for other energy bands for which the emission represent a small contribution to the total blazar emission (e.g. optical, X-rays), flux distributions close to log-normals (or double log-normals) have been found (e.g. Giebels \& Degrange 2009, Kushwaha et al. 2016, Kapanadze et al. 2020), suggesting that at these frequencies the variations are driven by different dynamical processes.

We note that our scenario is qualitatively different from self-organized criticality (SOC), the other possibility to obtain power law flux distributions mentioned by MSB19 (see e.g. Aschwanden et al. 2016). In fact SOC is based on the assumption that the system is continuously driven by an external energy source toward a critical threshold at which a rapid, non-linear phase is triggered, when the accumulated energy is released in an explosive fashion. In our framework, instead, the system always tries keep an equilibrium state and the dynamics is regulated by small perturbations continuously occurring in the structure. In this context, a possible difference that in principle can be used to distinguish between the two scenarios is the shape of the flares: while for SOC one expects a fast exponential grow followed by a slower decay (e.g. Aschwanden et al. 2016), more symmetric flares are expected in our scenario, because of the tendency of the drift term to keep the system in equilibrium. A more detailed comparison between the two scenarios, although interesting, is beyond the aim of this paper.

Finally we would like to remark that, although we motivated our SDE with a specific astrophysical framework, the same expression could also be applied to different scenarios. For instance, we can envisage an alternative scheme in which variability arises from processes related to the jet dynamics. In this scenario the drift could describe the tendency of the jet to keep a given radiative efficiency while the random term could account for the underlying perturbations.

\section*{Acknowledgments}
We thank the referee, G. V. Bicknell, for useful comments that helped us to improve the presentation. We thank Manuel Meyer and collaborators for providing us the files for the blazar light curves and the flux distributions used in the paper. We are grateful to G. Ghisellini for discussions and to G. Ghirlanda for useful suggestions. This work received contribution from the grant INAF Main Stream project "High-energy extragalactic astrophysics: toward the Cherenkov Telescope Array".

\section*{Data availability}

Data available on request.

\appendix

\section{Analytical solution of the stationary Fokker-Planck equation}
The stationary case of the Fokker-Planck equation (\ref{eq:fokker}) specialized to our case reads 
\begin{equation}
\frac{d}{d X}[\theta(\mu-X)p(X)]-\frac{d^2}{d X^2}\left[ \frac{\sigma^2 X^2}{2} p(X)\right]=0~,
\label{eq:fokkerApp1}
\end{equation}
which can be rewritten as
\begin{equation}
\frac{d}{d X} \left\{ \theta(\mu-X)p(X)-\frac{d}{d X}\left[ \frac{\sigma^2 X^2}{2} p(X)\right]\right\}=0~,
\label{eq:fokkerApp1a}
\end{equation}
by exploiting the linearity of the derivative operator. Now, we observe that the quantity in braces must be constant with respect to the variable $X$. Thus, from Eq. (\ref{eq:fokkerApp1a}) we obtain
\begin{equation}
\theta(\mu-X)p(X)-\frac{d}{d X}\left[ \frac{\sigma^2 X^2}{2} p(X)\right]=C~,
\label{eq:fokkerApp1b}
\end{equation}
where $C$ is a generic constant. By using the general method to solve the ordinary differential equations of the first order, we obtain the general solution
\begin{equation}
p(X)=\frac{e^{-\lambda\mu/X}}{X^{\lambda+2}}\left[k_1+k_2 \Gamma\left(-1-\lambda,-\frac{\lambda\mu}{X} \right) \right]~,
\label{eq:fokkerApp2}
\end{equation}
where $\Gamma(.,.)$ is the upper incomplete gamma function and $\lambda \equiv 2\theta/\sigma^2$, while $k_1$ and $k_2$ are two constants which must be determined by boundary conditions. For simplicity we have redefined $k_2 \equiv C(-\lambda\mu)^{\lambda+1}$ since it represents a generic constant. The two conditions we impose in order to find $k_1$ and $k_2$ are
\begin{equation}
\lim_{X \to \infty} p(X)=0~; \ \ \ \ \ \ \ \ \int_0^{\infty}dX \, p(X)=1~,
\label{eq:fokkerApp3}
\end{equation}
which express the conditions to have a vanishing probability at extremely high fluxes and the total probability to be unitary, respectively. However, the first condition is satisfied for all values of $k_1$ and $k_2$. Nevertheless, it is possible to infer that, for physically consistent values of the parameters ($\lambda,\mu>0$), the $\Gamma$ function produces complex values, so that the only possibility to have a real-valued $p(X)$ is to take $k_2=0$. As a result, Eq. (\ref{eq:fokkerApp2}) simplifies to
\begin{equation}
p(X)=k_1\frac{e^{-\lambda\mu/X}}{X^{\lambda+2}}~.
\label{eq:fokkerApp4}
\end{equation}
Now, by imposing the second condition (\ref{eq:fokkerApp3}), we obtain
\begin{equation}
k_1=\frac{ (\lambda \mu)^{1 + \lambda}}{\Gamma(1+\lambda)}~,
\label{eq:fokkerApp5}
\end{equation}
where $\Gamma$ is now the ordinary gamma function. Thus, the solution of Eq. (\ref{eq:fokkerApp1}) with physically consistent boundary conditions and values of the parameters $\lambda$ and $\mu$ reads
\begin{equation}
p(X)=\frac{ (\lambda \mu)^{1 + \lambda}}{\Gamma(1+\lambda)}\frac{e^{-\lambda\mu/X}}{X^{\lambda+2}}~.
\label{eq:fokkerApp6}
\end{equation}

\label{lastpage}

\end{document}